\documentstyle[manuscript,prl,aps]{revtex}
\begin{document}
\title{Mapping of the $B_N$-type Calogero-Sutherland-Moser
system to decoupled Harmonic Oscillators}
\author{N. Gurappa and Prasanta. K. Panigrahi}
\address{School of Physics \\
University of Hyderabad,\\
Hyderabad, Andhra Pradesh,\\
500 046 INDIA}
\maketitle

\begin{abstract} 
The $B_N$-type Calogero-Sutherland-Moser system in one-dimension is shown to
be equivalent to a set of decoupled oscillators by a similarity
transformation. This result is used to show the connection of the $A_N$ and
$B_N$ type models and explain the degeneracy structure of the later. We
identify the commuting constants of motion  and the generators of a {\it
linear} $W_\infty$ algebra associated with the $B_N$ system.
\end{abstract}
\vfill
$\dagger$ E-mail: panisp@uohyd.ernet.in
\draft
\pacs{}
\newpage
In the recent literature, there is an increasing interest in the quantum
mechanical, exactly solvable $N$-particle systems in one-dimension [1-4].
All these models are characterised by non-trivial, long-range interactions
between particles and Jastrow-type correlated many-body wave functions.
Interestingly, they have been found relevant for the description of universal
properties of various physical systems. These include spin chains [5],
quantum Hall effect [6], universal conductance fluctuations in mesoscopic
systems [7], two-dimensional gravity [8], gauge theories [9] and random
matrices [2,10].

In a recent paper [11], the present authors have
constructed a similarity transformation which maps the $A_N$-type
Calogero-Sutherland (CS) model [1,2], having pair-wise inverse-square and
harmonic interactions, to decoupled harmonic oscillators. Starting from the
symmetrised eigenfunctions of $N$ free harmonic oscillators, the wave
functions of the CS model can then be constructed explicitly. In general, the
Hamiltonians which can be brought through a suitable transformation to the
generalised form $\tilde H = \sum_i x_i \frac{\partial}{\partial x_i} + c +
\hat F$ can also be mapped to $\sum_i x_i \frac{\partial}{\partial x_i}  + c$
by a similarity transformation. The operator that accomplishes this is given
by $\exp\{- d^{-1} \hat F\}$; where, $\hat F$ is any homogeneous function of 
$\frac{\partial}{\partial x_i}$ and $x_i$ with degree $d$, and $c$ is a
constant. For the normalizability of the wave functions, one needs to check
that the action of $\exp\{- d^{-1} \hat F\}$ on an appropriate linear
combination of the eigenstates of $\sum_i x_i \frac{\partial}{\partial x_i}$
yields a polynomial solution. 

In this letter, the $B_N$-type Calogero-Sutherland-Moser (CSM) model [4] is
diagonalised using the above procedure. We (i) prove the equivalence of
$B_N$-type model to decoupled oscillators, (ii) construct the complete set of
eigenfunctions for this model from those of harmonic oscillators, (iii) show
the existence of $N$ independent commuting constants of motion and a {\it
linear} $W_\infty$ algebra. We also point out the connection of $A_N$ and
$B_N$-type models and explain the origin of degeneracies.

The CSM Hamiltonian (in the units $\hbar = m = \omega = 1$) is given by
\begin{equation}
H = - \frac{1}{2} \sum_{i=1}^N \partial_i^2
+ \frac {1}{2} \sum_{i=1}^N x_i^2 + \frac {1}{2} g^2
\sum_{{i,j=1}\atop {i\ne j}}^N \left\{\frac {1}{(x_i - x_j)^2} + 
\frac {1}{(x_i + x_j)^2}\right\} + \frac{1}{2} g_1^2
\sum_{i=1}^N \frac{1}{x_i^2} \qquad,
\end{equation}
where, $\partial_i \equiv \partial/\partial x_i$; $g^2$ and $g_1^2$ are
coupling constants. Notice that, the above Hamiltonian is not translational
invariant unlike the $A_N$-type model. Since, the ground-state wave function
of $H$, when the system is quantised as bosons, is given by
\begin{equation}
\psi_0 = \prod_{1\le {j < {k\le N}}}|x_i - x_j|^\lambda |x_i + x_j|^\lambda 
\prod_k^N |x_k|^{\lambda_1} \exp\{- \frac{1}{2}\sum_i x_i^2\}\qquad,
\end{equation}
one can make the following similarity transformation on the Hamiltonian
\begin{equation}
\tilde H \equiv \psi_0^{-1} H \psi_0 = \sum_i x_i \partial_i + E_0 + \hat F
\qquad.
\end{equation}
Here,
$$\hat F \equiv - \left(\frac{1}{2} \sum_i \partial_i^2 
+ \lambda \sum_{i<j} \frac {1}{(x_i^2 - x_j^2)} (x_i \partial_i - x_j
\partial_j)  + \lambda_1 \sum_i \frac{1}{x_i} \partial_i\right)\qquad,$$ 
$g^2 = \lambda (\lambda - 1)$, $g_1^2 = \lambda_1 (\lambda_1 - 1)$ and $E_0 =
N (\frac{1}{2} + (N-1) \lambda + \lambda_1)$ is the ground-state energy. 
The eigenfunctions of $\tilde H$ must be totally symmetric with respect to
the exchange of any two particle coordinates; the bosonic or the fermionic
nature of the wave function being contained in $\psi_0$.
One can easily establish the following commutation relation;
\begin{equation}
[\sum_i x_i \partial_i\,\,,\,\,\exp\{\hat F/2\}] = - \hat F \exp\{\hat
F/2\}\qquad. 
\end{equation}
Making use of (4) in (3), one gets
\begin{equation}
\exp\{- \hat F/2\} \tilde H \exp\{\hat F/2\} = \sum_i x_i \partial_i + E_0
\qquad. 
\end{equation}
Hence, the similarity transformation by the operator $\hat S \equiv \psi_0
\exp\{\hat F/2\}$ diagonalises $H$ {\it i.e.}, ${\hat S}^{-1} H \hat S =
\sum_i x_i \partial_i + E_0$. Furthermore, the following similarity
transformation on (5) makes the connection of $B_N$-type model with the
decoupled oscillators explicit:
\begin{equation}
G \hat E {\hat S}^{-1} H \hat S {\hat E}^{-1} G^{-1} =
-\frac{1}{2} \sum_i \frac{\partial^2}{\partial x_i^2} + \frac{1}{2} \sum_i
x_i^2 + (E_0 - \frac{1}{2} N) \qquad,
\end{equation}
where, $\hat E \equiv \exp\{-\frac{1}{4}\sum_i \frac{\partial^2} {\partial
x_i^2}\}$ and $G \equiv \exp\{- \frac{1}{2}\sum_i x_i^2\}$. This is one of the
main results of this letter.

At this stage, it is interesting to note that, (5) can also be made
equivalent to the Calogero-Sutherland model. This is achieved by
first defining an operator $\hat T \equiv Z G \exp\{- \hat A/2\}$, where $Z
\equiv \prod_{i<j} |x_i -  x_j|^\alpha$ and $\hat A \equiv \frac{1}{2} \sum_i
\partial_i^2 + \alpha \sum_{i<j} \frac{1}{(x_i - x_j)} (\partial_i -
\partial_j)$: 
\begin{equation}
\hat T \left(\sum_i x_i \partial_i + E_0\right) {\hat T}^{-1} = - \frac{1}{2}
\sum_i \partial_i^2 + \frac{1}{2} \sum_i x_i^2 +  \sum_{i<j} \frac{\alpha
(\alpha - 1)}{(x_i - x_j)^2} + (E_0 - E_0^\prime)\qquad,
\end{equation}
here, $E_0^\prime = \frac{N}{2}[1 + \alpha (N - 1)]$ is the ground-state
energy of the CS model. In other words, the operator $\hat S {\hat T}^{-1}$
maps CSM to CS model. 

From (5), one can also obtain the creation and annihilation operators as $a_i^+
= \hat S x_i {\hat S}^{-1}$ and  $a_i^- = \hat S
\partial_i {\hat S}^{-1}$: the symmetrised form of the operators $K_i^+ =
\frac{1}{2} {a_i^+}^2$  acts on the ground-state obtained from $a_i^- |0>
= 0; i = 1, 2,\cdots, N$ and creates the eigenstates of this $B_N$-type
model.  In terms of the creation and annihilation operators, the CSM
Hamiltonian can be written as $H = \sum_i H_i + (E_0 - N/2)$, where $H_i
\equiv a_i^+ a_i^- + \frac{1}{2}$. It can be easily checked that, the
commutation relations $[K_i^-\,\,,\,\,K_j^+] = \delta_{ij} H_i$ and
$[H_i\,\,,\,\,K_j^{\pm}] = \pm 2 \delta_{ij} K_i^{\pm}$ generate $N$
copies of $SU(1,1)$ algebra; here, $K_i^- = \frac{1}{2} {a_i^-}^2$. 
It is worth noticing here that these, $N$, commuting $SU(1,1)$ generators act
on the even sector of the harmonic oscillator basis and hence generate a
complete set of eigenfunctions. In this respect, it should be noted that, the
eigenfunctions of the CS model is made up of both the even and odd sectors of
the oscillator basis [11].

One can also define $<<0|{S}_n(\{\frac{1}{2}{a_i^-}^2\}) = <<n|$ and
${S}_n(\{\frac{1}{2}{a_i^+}^2\})|0> = |n>$ as the bra and ket vectors;
${S}_n$ is a symmetric homogeneous function of degree $n$ and
$<<0|\frac{1}{2}{a_i^+}^2 = \frac{1}{2}{a_i^-}^2 |0> = 0$. Since all the $N$ 
$SU(1,1)$ algebras are decoupled, the inner product between these bra and ket
vectors proves that any ket $|n>$, with a given partition of $n$, is
orthogonal to all the bra vectors, with different $n$ and also to those with
different partitions of the same $n$. The normalisation for any state $|n>$
can also be found out from the ground state normalisation, which is known
[13]. 

We can also make use of (5) for the explicit construction of the
eigenfunctions of (1). One 
notices that an arbitrary homogeneous function is an eigenfunction of $\sum_i
x_i \partial_i$; however, only the homogeneous symmetric functions of the
square of the particle co-ordinates are the ones, on which the action of
$\exp\{\hat F/2\}$ gives a polynomial solution. There are several related
basis sets available for these functions, viz., Schur functions, monomial
symmetric functions, complete symmetric functions and Jack polynomials[12].
Jack polynomials provide one special 
choice for this basis set which has been studied recently by T.H. Baker and
P.J. Forrester [13]. It is also worth noting that if $\lambda = 1$, then (5)
reduces to the differential equation, in $y_i = x_i^2$ variables for the
multivartiate Lagurre polynomials [14]. This is the basic reason why the
action of the operator $\exp\{\hat F/2\}$ on the homogeneous symmetric
functions of the square of the particle co-ordinates gives a polynomial
solution. For the sake of illustration, we choose here the power sum basis
$P_l\equiv \sum_i {(x_i^2)}^l$. The eigenfunctions and the energy eigenvalues
are respectively given by 
\begin{equation}
\psi_n = \psi_0 \left[\exp\{\hat F/2\} \prod_{l=1}^N P_l^{n_l}\right] \qquad,
\end{equation}
and $E_n = 2 \sum_l^N l n_l + E_0$; $n = \sum_l l n_l$. From this it is clear
that, $\hat S {\hat T}^{-1}$ maps all eigenfunctions of $H$ to
the even sector of CS model (or to that of harmonic oscillators, if $\alpha =
0$ or $1$).

The quantum integrability of the CSM model and the identification of the
constants of motion become transparent after establishing its equivalence to
free oscillators. It is easy to verify that $[H\,\,,\,\,H_k] =
[H_i\,\,,\,\,H_j] = 0$ for $i,j,k = 1,2,\cdots, N$. Therefore, the set
$\{H_1, H_2,\cdots, H_N\}$ provides the $N$ conserved quantities. One can
construct linearly independent symmetric conserved quantities from the
elementary symmetric polynomials, since they form a complete set:
\begin{equation}
I_1 = \sum_{1 \le i \le N} H_i \,\,, I_2 = \sum_{1 \le i < j \le N} H_i
H_j \,\,, I_3 = \sum_{1 \le i < j < k \le N} H_i H_j H_k,
\cdots, I_N = \prod_{i=1}^N H_i.
\end{equation}
Here, we would like to point out that, the present proof is entirely
different from earlier works [15]. 

With the generators of the above mentioned $SU(1,1)$ algebras of CSM, one can
define a linear $W_\infty$ algebra for which there exist several basis sets
[16]. 

In conclusion, we mapped the $B_N$-type Calogero-Sutherland-Moser model to
decoupled oscillators and algebraically constructed the complete
set of eigenfunctions, including the degenerate ones, from the symmetrised
eigenstates of the harmonic oscillators. We also showed the existence of $N$
independent commuting constants of motion, a {\it linear} $W_\infty$ algebra
for this system and the connection it has with $A_N$-type model. The
application of this method to other one and higher-dimensional [17] models
can also be carried out in an analogous manner.

The authors are grateful to Profs. V. Srinivasan and S. Chaturvedi
for many illuminating discussions. P.K.P would like to acknowledge useful
discussions with Prof. A. Khare and Dr. M. Sivakumar. N.G thanks E. Harikumar
for useful comments and U.G.C (India) for financial support through the S.R.F
scheme.

\end{document}